\documentclass[a4paper]{jpconf}
\usepackage{graphicx}
\usepackage{lineno}
\usepackage{hyperref}
\usepackage{float}
\usepackage{lineno}

\begin{document}

\title{Techniques and tools for measuring energy efficiency of scientific software applications}

\author{David Abdurachmanov$^1$, Peter Elmer$^2$, Giulio Eulisse$^3$, Robert Knight$^4$, Tapio Niemi$^5$, Jukka K. Nurminen$^6$, Filip Nyback$^6$, Gon\c{c}alo Pestana$^5$ $^6$, Zhonghong Ou$^6$, Kashif Khan $^5$ $^6$}

\address{$^1$ Digital Science and Computing Center, Faculty of Mathematics and Informatics, Vilnius University, Vilnius, Lithuania}
\address{$^2$ Department of Physics, Princeton University, Princeton, NJ 08540, USA}
\address{$^3$ Fermilab, Batavia, IL 60510, USA}
\address{$^4$ Research Computing, Office of Information Technology, Princeton University, Princeton, New Jersey 08540, USA}
\address{$^5$ Helsinki Institute of Physics, PO Box 64, FI-00014, Helsinki, Finland }
\address{$^6$ Aalto University, PO Box 11100, 00076 Aalto, Finland}

\ead{goncalo.pestana@aalto.fi}

\begin{abstract}
The scale of scientific High Performance Computing (HPC) and High 
Throughput Computing (HTC) has increased significantly in recent years,
and is becoming sensitive to total energy use and cost.
Energy-efficiency
has thus become an important concern in scientific fields such as High
Energy Physics (HEP). There has been a growing interest in utilizing
alternate architectures, such as low power ARM processors, to replace 
traditional
Intel x86 architectures. Nevertheless, even though such solutions
have been successfully used in mobile applications with low I/O and
memory demands, it is unclear if they are suitable and more
energy-efficient in the scientific computing environment. Furthermore,
there is a lack of tools and experience to derive and compare power 
consumption between the architectures for various workloads, and 
eventually to support software optimizations for energy efficiency.
To that end, we have performed several physical and software-based
measurements of workloads from HEP applications running on ARM and Intel
architectures, and compare their power consumption and performance.
We leverage several profiling tools (both in hardware and software)
to extract different characteristics of the power use. 
We report the results of these measurements and
the experience gained in developing a set of measurement techniques
and profiling tools to accurately assess the power consumption for
scientific workloads. 
\end{abstract}

\section{Introduction}

The Large Hadron Collider (LHC)~\cite{LHCPAPER} at the European
Laboratory for Particle Physics (CERN) in Geneva, Switzerland, is
an example of a scientific project whose computing resource requirements are
larger that those likely to provided in a single
computer center. Data processing and storage are distributed
across the Worldwide LHC Computing Grid (WLCG)~\cite{WLHC}, which
uses resources from 160 computer centers in 35 countries.
Such computational resources have enabled the
CMS~\cite{CMSDET} and ATLAS~\cite{ATLAS} experiments 
to discover the Higgs Boson~\cite{CMSHIGGS,
ATLASHIGGS}, for example. The WLHC requires
a massive amount of computational resources (250,000 x86 cores in
2012) and, proportionally,
energy. 
In the future, with planned increases to the LHC luminosity~\cite{HLLHC},
the dataset size will increase by 2-3 orders of magnitude, presenting even 
more challenges in terms of energy consumption. 


In order to find and develop better solutions for improving energy
efficiency in High Energy Physics (HEP) computing, it is important to understand
how energy is used by the HEP systems themselves. We describe several
tools and techniques that facilitate researchers to reach that goal.

As energy efficiency becomes a concern, new solutions have been
considered to develop energy efficient systems. One potential
solution is to replace the traditional Intel x86 architectures by
low power architectures such as ARM. A comparison of the energy
efficiency between ARMv7 and x86 Intel architecture is conducted
in this article. The experiments use CMS workloads and rely on the
techniques and tools described earlier to perform the measurements.

This article is structured as following. Firstly, we describe where
is energy consumed in a HTC system and outline some of
the tools and techniques available to measure and monitor energy
consumption on HTC systems (Section 2). Secondly, we present the results of a
comparison between ARMv7 and Intel Xeon architecture using CMS
workloads (Section 3). Finally, we present IgProf, a general purpose, open source
 application performance profile. In addition, we describe its recent added 
energy profiling features and 64-bit ARM support.

\section{Tools and techniques for energy measurement}

When optimizing power usage, there are two granularities at which
one can look at a computing system. The coarser granularity
takes into account the behavior of the whole node (or
some of its passive parts, e.g.\ the transformer) as part of a rack
in a datacenter. This is usually investigated when
engineering and optimizing computing centers. Alternatively,
a more detailed approach is to
look into the components which make up the active parts of a
node, in particular the CPU and its memory subsystem since these
are responsible for a sizeable fraction of the consumed power.
They are also the place where the largest gains in terms of efficiency 
can be obtained through optimizations in the software.

If one is simply interested in the coarse power consumption by node,
external probing devices can be used: monitoring interfaces
of the rack power distribution units, plugin meters and non-invasive
clamp meters (allowing measurement of the
current pulled by the system by induction without making physical
contact with it). They differ mostly in terms of flexibility.
Their accuracy is typically a few percent for power, whereas their time
resolution is in the order of seconds. This is more than enough
to optimize electrical layout of the datacenters or to provide a
baseline for more detailed studies.

\begin{figure}[tbp]
\centering
\includegraphics[width=70mm]{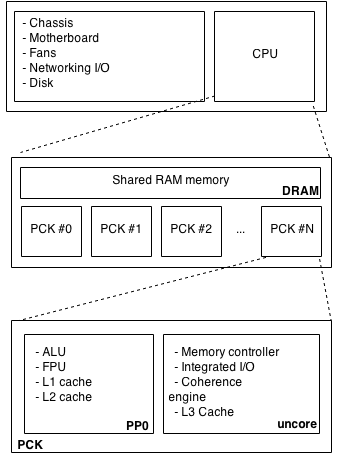}
\caption{Components that contribute for power consumption in HPC}
\label{fig:power-consumption-model}
\end{figure}

A alternative approach takes into account the internal structure of a
computing element of an HTC system, as shown in figure 
~\ref{fig:power-consumption-model}. Nowadays, every board manufacturer
provides on-board chips which monitor energy consumption of
different components of the system. These
allow energy measurements of fine grained detail, as it is possible
to individually monitor energy consumption of components such as
the CPU, its memory subsystem, and others. An example of this chip
monitors is the Texas Instruments TI INA231~\cite{TIINA231} current-churn
and power monitor which is found on the ARMv7 developer board which
we used for our studies. It is quite common in the industry.
Compared to external methods, these on-board components provide
high accuracy and reasonably high precision measurements (millisecond
level).

A special and slightly different case of these on-board monitors
is a new technology called Running Average Power Limit (RAPL),
provided by Intel beginning from the Sandy Bridge family of processors.

Contrary to other solutions, which are implemented as discrete
chips, RAPL is embedded as part of the CPU package itself and
provides information on the CPUs own subsystems. In particular RAPL
provides data for three different domains: \textbf{package} (pck),
which measures energy consumed by the system's sockets, \textbf{power
plane 0} (pp0), which measures energy consumed by the CPU core(s),
and \textbf{dram}, which accounts for the sum of energy consumed
by memory in a given socket, therefore excluding the on-core
caches~\cite{INTELMAN}. As for the discrete components case, the
timing resolution of measurements is in the millisecond range~\cite{RAPL1}.
This is fine enough to permit exploiting such data to build an energy
consumption sampling profiler for applications, similar to how performance
sampling profilers work (see section~\ref{sec:sampling}).
Finally, in addition to power
monitoring of the sockets, RAPL can limit the power consumed by the
different domains. This feature, usually referred as power capping,
allows the user to define the average power consumption limit of a
domain in a defined time window and allows more accurate independent
measurements of the non limited components.

\section{Power efficiency measurements with x86-64 and ARMv7}

In this section, we demonstrate the potential of some of the
tools we previously described. To that end, we perform
several measurements of workloads from CERN, running on different
architectures. The workloads used in the experiment run on top of
Intel x86-64 architecture, traditionally used in HTC and data centers
and 32 bit ARMv7 architectures (for similar studies for 64bit ARMv8
and Xeon Phi, please refer to~\cite{ABD2014}). The ARM architecture,
initially developed for mobile devices, has been
considered~\cite{ACAT13ARM, CHEP13ARMPHI} as a potential alternative
to Intel in HTC, given its energy efficient computing. We also
present a brief comparison between ARM and Intel architectures from
the energy consumption perspective, based on the results obtained.

\subsection{Tools and techniques}

\begin{figure}[ht!]
\centering
\includegraphics[width=150mm]{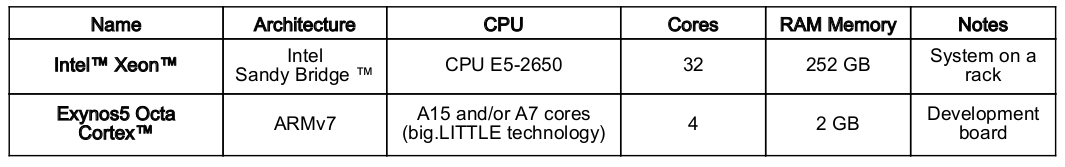}
\caption{Machine's specifications}
\label{figure:machine-specs}
\end{figure}

For the Intel architecture, we used the RAPL technology to perform measurements
of the energy consumed by the package, DRAM and cores 
(figure~\ref{fig:power-consumption-model}).
The external measurements for the baseline were performed using a
rack PDU, which provides an online API to gather the energy consumed
by the system on the rack at a sampling rate of 1 second.
For the ARM board, we used the Texas Instrument power monitor
chip TI INA231 which allows reading of the energy consumed by the
cores and dram at a sampling rate of microseconds. The chip was
embedded in the board from the vendor. For the external measurements,
we used an external plug-in power monitor with a computer interface
for gathering and storing the results.
In both cases we read the data as it was exposed to the system via
the sysfs / devfs knobs.
The machine specifications can be seen in figure~\ref{figure:machine-specs}.

\subsection{Experiment setup}

\begin{figure}[tbp]
\centering
\includegraphics[width=170mm]{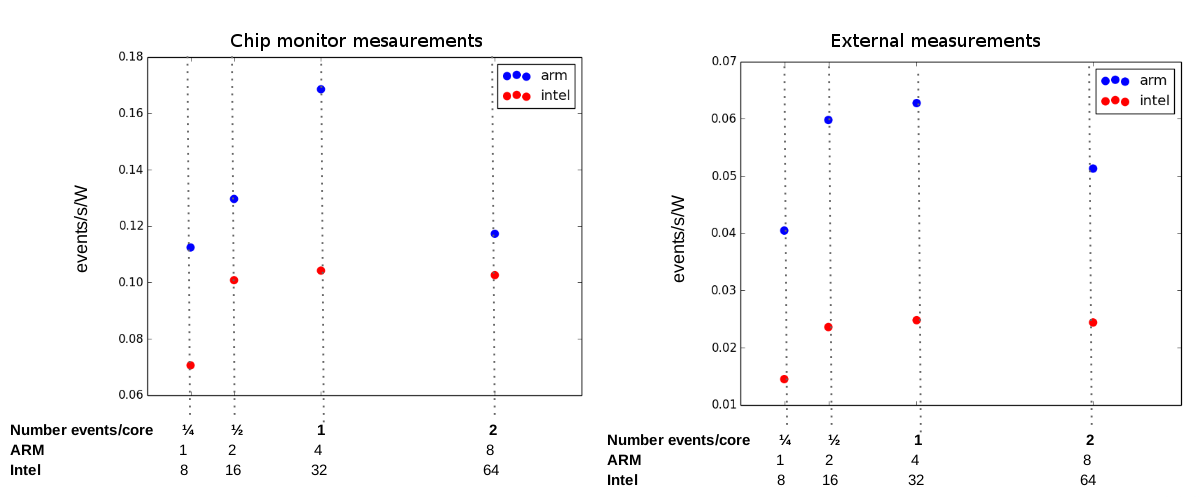}
\caption{Chip monitor and external measurements results. The results are shown
according to the relation events per number of cores of each machine and their
absolute number of cores. }
\label{fig:parfull-cms-benchmark}
\end{figure}

The workload used for the experiments was ParFullCMS, a multi-threaded
Geant4~\cite{GEANT4} benchmark application which uses a complex CMS
geometry for its simulation. Using ParFullCMS, we ran simulation
tasks on both the Intel and ARM machines 
(figure~\ref{fig:parfull-cms-benchmark}).  The workflow was run several times
with different number of threads in each machine. The number of
threads run in each experiment is chosen according to the number of
the cores of the machines.

\subsection{Analysis}

As expected, the ARMv7 architecture shows encouraging results from
the energy efficiency perspective than Intel in all the experiments
performed. Also as expected, both architectures do not perform
better when overcommitted (more threads than the physical number
of cores).  Notice ARM results when overcommitted
(8 threads) are much worse then Intel ones. This is due to the relatively 
modest amount of available DRAM (figure~\ref{figure:machine-specs}), causing 
the machine started swapping,
greatly affecting performance. While this was expected, since the
ARM system used is just a development board for mobile applications,
this is a clear indication that when doing a final assessment of
power efficiency for an architecture, one needs to have a full 
server-grade system in order to make a proper comparison.

\section{Profiling for power efficiency}
\label{sec:sampling}

The hardware components described above provide measurements that
are related to the full set of processes running on the machine.
For the simple case where only a single benchmark application is running, 
these can be used to make comparisons between architectures.
A further step is to try to see if there is a way to map the energy 
consumption measurements to functions and methods within an
executing process. Such a mapping would allow for optimizations of
the software itself.
This kind of mapping can be done in two different ways, which we call
{\it instrumentation} and {\it sampling profiling}.

In the {\it instrumentation} case, effective readings of profiled
quantities (e.g.\ energy consumption), or quantities correlated with
them (e.g.\ CPU power state transitions) are done at the beginning
and the end of a profiled task and the difference between the two
is used to estimate average power consumption over that period of
time. By bookkeeping starting and stop values for monitored tasks
one can get a fairly complete picture of what is happening to the
system,  provided the measured interval is large compared to the
temporal precision of the measure being done. This is both to avoid
a large error on the average estimation and to reduce performance
overhead due to the measure itself.

{\it Sampling profiling}, on the other hand, has a different approach where
a given quantity is sampled regularly and at each sample the measured
quantity is accumulated until it overflows a user provided limit.
When this happens the profiler increments a counter for the process /
function being executed in that precise moment. Assuming that the
distribution of where time is spent in a system is constant
over time (which is typically true for large data processing tasks),
such a sampling algorithm converges to the actual distribution of the measured
quantity. The advantage of this approach is that the fidelity of
the measurement to first approximation depends only on the
number of samples made, regardless of the error on the profiled
quantity. This also allows minimizing the performance overhead by tuning the 
sampling period to be much larger than the measurement itself.

\texttt{IgProf} is a general purpose, open source application performance 
profiler. It was developed in HEP, but it is capable of profiling all 
types of software applications.
The profiler has been available on the x86 and x86-64 platforms
since many years~\cite{igprofchep04, igprof-web},
and recently we have also ported it to ARMv7 and ARMv8. 
Moreover we have now added a statistical sampling energy
profiling module which provides function level energy cost
distribution~\cite{weaver12}. Such a module uses the PAPI library
to read energy measurements from the RAPL interface previously
described.

\begin{figure}[tbp]
  \begin{center}
    \includegraphics{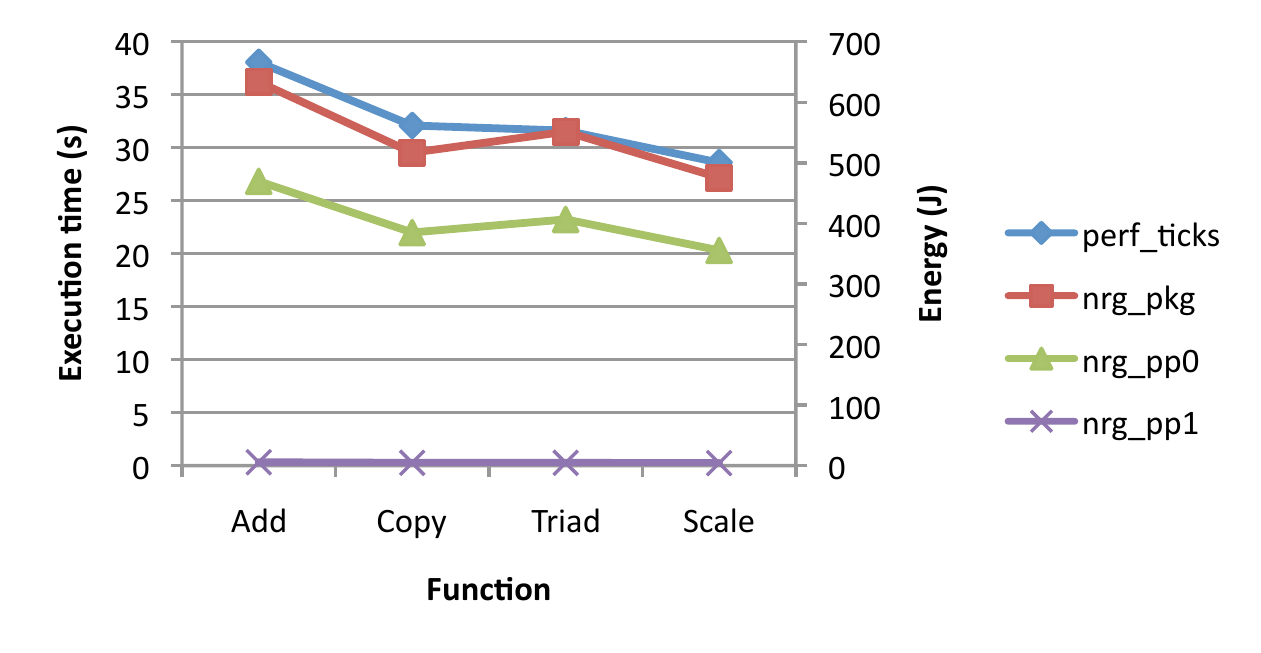}
  \end{center}
  \vspace{-20pt}
  \caption{The results of performance and energy profiling of the \texttt{STREAM} tool.}
  \label{fig:stream-pp-np}
\end{figure}

To illustrate the new module, we use it to profile the memory
benchmark \texttt{STREAM}~\cite{stream-web}. Figure~\ref{fig:stream-pp-np}
compares the results from performance and energy profiling of the
benchmarking tool. The X-axis describes the four main functions
contributing to the execution time and energy consumption of the
stream tool: Add, Copy, Triad and Scale. The left scale of the
Y-axis and the perf\_ticks series describe the execution time spent
in each function, whereas the right scale of the Y-axis and the
nrg\_pkg, nrg\_pp0 and nrg\_pp1 series describe the amount of energy
spent in each function. The energy consumption of the processor
package domain and the power plane 0 (describing the CPU cores)
seem to follow the time spent in the functions, whereas the energy
consumption of power plane 1 seems to be fairly constant to zero
(describing the unused GPU) .

As we would expect from a simple benchmark, the profiling results of
a simple single-threaded application shows a correlation between
the execution time and the energy spent in a function.
While the energy profiling module is now fully functional within 
\texttt{IgProf}, further work needs to be done to tune the measurements 
and to gain experience with how to use the profiles obtained for
more complex applications.

\section{Conclusions}

Energy efficiency has become a major concern for HTC, given the large
amount of computing resources - and thus energy - that recent experiments
require. LHC computing is a prime example of the need for energy
efficient facilities, given its present requirements and costs
constraints. The need for energy efficiency drives an interest
in accurately evaluating the different components of a HTC system
to understand how and where energy is consumed
and improve the overall efficiency. However, HTC systems are complex
and composed of different components. In this paper we have presented
a number of techniques and tools that provide insight into how and
where energy is consumed from different perspectives and granularities.
In addition, \texttt{IgProf}, an open source profiling tool, has been
extended
to run on 64-bit ARM and to provide function-level energy profiling capabilities. 
Using these tools and techniques we have also reported studies done to
compare the energy performance of x86-64 and ARMv7 processors, confirming
the potential of ARMv7 for efficient HTC systems should server grade
systems be built around such chips.

\section*{Acknowledgements}
This work was partially supported by the National Science Foundation,
under Cooperative Agreement PHY-1120138, and by the U.S. Department
of Energy. ARMv8 and energy profiling support in IgProf was also 
supported by Google Summer of Code (GSoC 2014).

\section*{References}

\bibliographystyle{unsrt}
\bibliography{acat2014-energy-efficiency}

\end{document}